\documentstyle[pra,aps]{revtex}
\begin{document}
\draft
\title{Normal modes of a vortex in a trapped
Bose-Einstein condensate}
\author{Anatoly A.~Svidzinsky and Alexander
L.~Fetter}
\address{Department of Physics, Stanford
University, Stanford, CA 94305-4060}
\date{\today}
\maketitle
\begin{abstract}
A hydrodynamic description  is used to study  the normal modes of a vortex
in a zero-temperature Bose-Einstein  condensate.  In the Thomas-Fermi (TF)
limit,
the circulating superfluid velocity far from the vortex
 core provides a small perturbation that splits  the originally
degenerate  normal modes of a vortex-free condensate.  The relative frequency
shifts are
 small in all cases considered (they vanish for the lowest dipole mode with
$|m|=1$), suggesting that the vortex is stable.  The Bogoliubov equations serve
to verify the existence of helical waves, similar to
   those of a vortex line in an unbounded weakly interacting Bose gas.
In the large-condensate (small-core) limit, the condensate wave function
reduces to that of
 a straight vortex in an unbounded condensate; the corresponding Bogoliubov
equations have no bound-state solutions that are uniform along the symmetry
axis and decay exponentially far from the vortex core.
\end{abstract}

\section{Introduction}

Recent experimental demonstrations of
 Bose-Einstein condensation in dilute low-temperature trapped
alkali gases \cite{And,Dav,Brad}   rapidly stimulated  detailed
theoretical \cite{Edw,Str,EDCB,Rup} and experimental \cite{Jin,MOM} studies of
the lowest-lying normal modes.  These latter experiments confirmed the
essential
applicability of the Bogoliubov approximation \cite{Bog},  which assumes that
most of the particles  remain in the self-consistent condensate for
temperatures
well below the Bose-Einstein condensation temperature.  Although Bogoliubov's
original work described only  a uniform system, the formalism was soon extended
to   a vortex line
 \cite{EPG,LPP} and other more general nonuniform  configurations \cite{AP}.

In a  harmonic trap with anisotropic frequencies $\omega_\alpha$
($\alpha = x$, $y$, and
$z$), an
ideal-Bose-gas condensate would have characteristic dimensions $d_\alpha
=\sqrt{\hbar/M\omega_\alpha}$, where $M$ is the atomic mass. As the number
$N$ of
particles grows, however, the repulsive interactions become important, and
  the condensate expands  from the noninteracting  mean oscillator
radius
$d_0= (d_xd_yd_z)^{1/3}$ to a larger mean radius
$R_0$.  For $R_0\gg d_0$, the  kinetic energy of a particle in the
condensate is  negligible relative to the trap potential energy and the
repulsive interparticle
potential energy, leading to the Thomas-Fermi (TF) approximation for the
condensate density
\cite{BP}.  This limit occurs when the condensate number $N_0$ is much larger
than the ratio $d_0/a$, where  $a$ is the effective
$s$-wave interatomic scattering length (typically \cite{Mewes} $d_0\approx$
a few
$\mu m$,
$a\approx$  a few  nm,  and the experimental value $N_0\gtrsim 10^6$
amply satisfies the  TF criterion  that
$N_0
\gg 10^3$).  In the TF regime, the expansion ratio is given by  $R_0/d_0
\approx 15 (N_0a/d_0)^{1/5}\gg 1$ \cite{BP}.

The normal modes of a nonuniform condensate can be described  with either the
pair
of quantum-mechanical quasiparticle amplitudes $u$ and $v$  that obey the
coupled Bogoliubov
equations
\cite{LPP,AP,deG}, or the equivalent  hydrodynamic amplitudes for the
fluctuations $n'$ in the particle  density and $\Phi'$ in the velocity
potential
\cite{Str,ALF,WG}.  In the TF limit, the latter description simplifies
considerably for the low-lying
normal modes, allowing accurate predictions of  lowest eigenfrequencies of an
axisymmetric trapped condensate
\cite{Str,Jin,MOM}.

In contrast, the creation and detection of a vortex  in a trapped Bose
condensate remains relatively unexplored.  The condensate density for
a vortex line on the axis of an axisymmetric harmonic potential has been
evaluated both numerically
\cite{Edw2,DS} and with the TF approximation
\cite{SS,DR1}.    Using the Bogoliubov equations in the TF limit,  Sinha
\cite{SS} studied the
  normal modes of a condensate with a vortex, showing that the presence of the
vortex causes only a small shift  of order
$(d_0/R_0)^2\ll 1$ for a subset of the eigenfrequencies.  In addition, a
detailed  numerical study has been carried out \cite{DBEC} for for the
specific trap configuration of Ref.~\cite{And} for
$Na/d_0\lesssim 50$
 (for the distinct but related example of  an annular condensate, see
Ref.~\cite{DR2} and references therein). Finally, Zambelli and
Stringari~\cite{ZS}
 have used sum rules to calculate the splitting of the lowest (quadrupole)
normal modes caused by the presence of a vortex.

 The present work uses the equivalent
hydrodynamic picture  to study the low-lying normal modes of a quantized
vortex line on the symmetry axis
of an axisymmetric harmonic trap.  In the TF limit, the vortex core radius
$\xi\sim d_0^2/R_0$ is small compared to both the mean oscillator length $d_0$
and the mean dimension
$R_0$ of the condensate.  The resulting hydrodynamic equations then yield a
perturbative treatment of the normal-mode eigenfrequencies and eigenfunctions
for the density fluctuations, in which the vortex's circulating velocity
splits
the degenerate unperturbed eigenfrequencies with azimuthal angular momentum
$\pm m$ by a relative amount of order
$\xi/R_0\sim (d_0/R_0)^2\ll 1$, largely confirming Sinha's conclusion \cite{SS}
based on the Bogoliubov equations (for $|m|=1$, however, we find additional
terms which ensure that the shift vanishes for the lowest dipole mode).  As
emphasized by Zambelli and Stringari~\cite{ZS}, this splitting provides a
sensitive test of the existence of a vortex.

 Since the vortex-free trapped
condensate is stable (both experimentally and theoretically), these results
suggest that the corresponding trapped vortex  is also stable at zero
temperature. In the presence of weak dissipation, however, a slightly
off-center vortex presumably will slowly spiral outward and eventually
disappear, as noted by Hess~\cite{Hess} and by Packard and Sanders~\cite{PS}
for superfluid ${}^4$He and by Rokhsar~\cite{DR1} for trapped condensates.

Section II reviews the basic hydrodynamic formalism, focusing on
a condensate containing a $q$-fold
quantized vortex  in an axisymmetric trap. For the normal modes $\propto
e^{im\phi}$, a general variational principle  demonstrates that the
frequencies are manifestly non-negative (and hence  stable) for
$m=0$ and for $m^2\ge 4q^2$ (Sec.~V gives a  more
detailed treatment for the remaining cases,    but only in the TF limit).
For a large condensate (the TF limit), Sec.~III makes use of the small
parameter $\xi/R_0$ to perform a perturbation expansion of the hydrodynamic
equations, yielding a splitting of the originally degenerate modes with $\pm
m$.  In Sec.~IV, the relative frequency shift of all low-lying normal modes is
evaluated analytically for a spherical trap,  whereas only a subset of the
modes allows a closed-form expression for a strictly cylindrical trap and for a
general axisymmetric but anisotropic trap.  Section V contains additional
considerations on the stability of a vortex line, based  on the full Bogoliubov
equations, including a demonstration that all normal modes with
$k=0$  have non-negative eigenfrequencies in the TF limit (and thus are
stable).

\section{Basic formalism}
Consider a trapped condensate characterized by the equilibrium condensate wave
function $\Psi$. It satisfies the Gross-Pitaevskii (GP) equation \cite{EPG,LPP}
\begin{equation}
\big(T  + V_{\rm tr} + g|\Psi|^2 -\mu\big)\Psi =
0,\label{GP1}
\end{equation}
where $T= -\hbar^2\nabla^2/2M$ is the kinetic-energy operator, $V_{\rm tr}$
is the external trap potential, $g = 4\pi a\hbar^2/M$ is the effective
interparticle interaction strength (expressed in terms of the $s$-wave
scattering length $a$, here taken as positive), and
$\mu$ is the chemical potential.
 It is convenient to write $\Psi = |\Psi|\,e^{iS}$, where $n_0 = |\Psi|^2$
is the
static condensate particle density
 and  ${\bf v}_0 =
(\hbar/ M)  \bbox{\nabla} S$ is the static condensate velocity.  These
quantities satisfy the static conservation law
$\bbox{\nabla\cdot}(n_0{\bf v}_0)= 0$,
which is the imaginary part of the GP equation.
Correspondingly, the real part
\begin{equation}
\mu =|\Psi|^{-1}T|\Psi| +\case{1}{2}Mv_0^2 + V_{\rm tr} + g|\Psi|^2,
\label{GP2}\end{equation}
  generalizes  Bernoulli's equation to include the quantum
(kinetic-energy)
contribution $|\Psi|^{-1}T|\Psi|$.

Hydrodynamics relies on the local density and velocity as the relevant
variables.  In the present context, the linearized  hydrodynamic equations
determine the  harmonic fluctuations in the particle density
$ n' e^{-i\omega t}$ and the velocity potential $
\Phi' e^{-i\omega t}$.  The corresponding (complex) normal-mode amplitudes
satisfy the  coupled equations
\cite{Str,ALF}
\begin{mathletters}
\label{hyd}
\begin{eqnarray}
i\omega n' &=& \bbox{\nabla\cdot}\big({\bf v}_0\,n'\big) +
\bbox{\nabla\cdot}\big( n_0\,\bbox{\nabla}\Phi'\big), \label{hyd1}\\
i\omega \Phi'& = &{\bf v}_0\bbox{\cdot\nabla}\Phi' +
\frac{g}{M} n' -
\frac{\hbar^2}{4M^2n_0}\bbox{\nabla\cdot}\left[n_0\bbox{\nabla }
\left(\frac{ n'}{n_0}\right)\right],\label{hyd2}
\end{eqnarray}
\end{mathletters}
both of which involve  the (real) static condensate density $n_0$ and
(real) condensate velocity
${\bf v}_0$.
These very general equations can be derived from either  the
linearized quantum-field equations or the  conservation of particles
and Bernoulli's equation for irrotational isentropic  compressible flow
\cite{ALF,WG}. The last term in Eq.~(\ref{hyd2}) corresponds to the kinetic
energy (quantum) pressure and is omitted in  the TF limit of a large
condensate. Although the resulting formalism is frequently known simply as
the ``hydrodynamic approach,''
 we prefer to use the term
``hydrodynamic equations" for the more complete Eqs. (\ref{hyd}).

The appropriate normalization of these amplitudes follows from that of the
 Bogoliubov amplitudes, which satisfy the condition  \cite{AP} $\int
dV\,\left(|u|^2-|v|^2\right) = 1$.  A straightforward comparison of
the linearized quantum-field equations in the presence of
condensate flow  \cite{ALF} shows that
\begin{mathletters}\label{n'Phi'}
\begin{eqnarray}
n'&=&\Psi^* u-\Psi v,\label{n'}\\
\Phi'&=&\frac{\hbar}{2Mi|\Psi|^2}(\Psi^* u+\Psi v).\label{Phi'}
\end{eqnarray}
\end{mathletters}
 As a result, we have
\begin{equation}
|u|^2-|v|^2=\frac{iM}{\hbar}\left(n'^*\Phi' -
\Phi'^*n'\right),
\end{equation}
so that the normalization of the hydrodynamic amplitudes requires
\begin{equation}\int dV\,i\left(n'^*\Phi' -
\Phi'^*n'\right)=\frac{\hbar}{M}.\label{hydnorm}
\end{equation}

Multiply Eq.~(\ref{hyd1}) by $\Phi'^*$,  multiply
Eq.~(\ref{hyd2}) by $n'^*$,   integrate the results over all space, and
subtract the first from the second.  A simple integration by parts yields
\begin{equation}
\omega\int dV \,i(n'^*\Phi'-\Phi'^*n') = \int dV\bigg[\left(n'^*{\bf
v}_0\bbox{\cdot\nabla}\Phi'+n'{\bf
v}_0\bbox{\cdot\nabla}\Phi'^*\right)+n_0\left|\bbox{\nabla}\Phi'\right|^2
+\frac{g}{M}\left|n'\right|^2+\frac{\hbar^2}{4M^2}n_0\left|\bbox{\nabla}\left(
\frac{n'}{n_0}\right)\right|^2\bigg]. \label{var1}
\end{equation}
In this way, the normal-mode frequency  is expressed as the ratio of two
manifestly real integrals, so that $\omega$ itself must be real [assuming
that the normalization integral Eq.~(\ref{hydnorm}) is nonzero].  In addition,
variation of Eq.~(\ref{var1})  with respect to each of the two hydrodynamic
variables
$n'^*$ and $\Phi'^*$ reproduces the dynamical equations (\ref{hyd}), so that
Eq.~(\ref{var1}) serves as a variational principle for the normal-mode
eigenfrequencies. Finally, Eq.~(\ref{hydnorm}) shows that  the left-hand side
here is just
$\hbar\omega/M$ for a properly normalized set of solutions.

We consider a quantized vortex with $q$ quanta of circulation on the
 symmetry axis of an axisymmetric trap with
\begin{equation}
V_{\rm tr} = \case{1}{2}M\big(\omega_\perp^2\rho^2 +
\omega_z^2z^2\big)=\case{1}{2}M\omega_\perp^2\big(\rho^2 +
\lambda^2z^2\big),\label{pot}
\end{equation}
where ($\rho$, $\phi$,  $z$) are the usual cylindrical polar coordinates, and
we define the anisotropy parameter $\lambda = \omega_z/\omega_\perp$. The
appropriate solution of the GP equation (\ref{GP1}) has the form
$\Psi_q=e^{iq\phi}\,|\Psi_q(\rho, z)|$, so that the condensate velocity can
be written ${\bf v}_0=V\hat\phi$, where
\begin{equation}
V=\frac{\hbar q}{M\rho}.
\end{equation}
In this case, the
linearized hydrodynamic equations (\ref{hyd1}) and (\ref{hyd2}) have
single-valued
solutions of the form
\begin{equation}
n'({\bf r}), \;\Phi'({\bf r})\propto \exp (im\phi),
\end{equation}
where $m$ is an integer and the corresponding amplitudes depend only on
$\rho$ and $z$.  Substitution into Eq.~(\ref{var1}) yields
\begin{eqnarray}
\frac{\hbar\omega}{M}
&=& \int
dV\bigg[\frac{mq}{\rho^2}\frac{\hbar}{M}\,i\left(n'^*\Phi'-n'\Phi'^*\right)
+
\frac{m^2}{\rho^2}\left(n_q|\Phi'|^2
+\frac{\hbar^2}{4M^2n_q}|n'|^2\right)\nonumber\\
& \quad+& n_q\left|\bbox{\nabla}\Phi'\right|^2
+\frac{g}{M}\left|n'\right|^2+\frac{\hbar^2}{4M^2}n_q\left|\bbox{\nabla}\left(
\frac{n'}{n_q}\right)\right|^2\bigg], \label{varprime}
\end{eqnarray}
where $n_q=|\Psi_q|^2$ is the (axisymmetric)  condensate density for a $q$-fold
vortex.

 Simple manipulation yields the equivalent form
\begin{eqnarray}
\frac{\hbar\omega}{M}
&=& \int
dV\bigg[\frac{1}{\rho^2n_q}\,\bigg|\,m\,\frac{\hbar}{2M}\,
n'+2q\,i\,n_q\,\Phi'\bigg|^2
+\frac{m^2-4q^2}{\rho^2}\,n_q\,|\Phi'|^2\nonumber\\ \qquad& \quad+&
n_q\left|\bbox{\nabla}\Phi'\right|^2
+\frac{g}{M}\left|n'\right|^2+\frac{\hbar^2}{4M^2}n_q\left|\bbox{\nabla}\left(
\frac{n'}{n_q}\right)\right|^2\bigg], \label{var2}
\end{eqnarray}
 Here, the right-hand side is manifestly non-negative  for
$m=0$ and for $m^2\ge 4q^2$, so that all such hydrodynamic modes of a
general
$q$-fold quantized vortex necessarily  have non-negative frequencies. As
discussed in Sec.~V, the quantum-mechanical ``grand-canonical'' hamiltonian
in the Bogoliubov approximation reduces to a set of uncoupled harmonic
oscillators for the different quasiparticle modes and is thus bounded from
below so long as none of the eigenfrequencies is negative.  Consequently,
the vortex is stable with respect to modes with $m=0$ and $m^2\ge 4q^2$. For
the remaining modes with $0<m^2<4q^2$, Eq.~(\ref{var2}) contains terms of both
signs that can in principle produce an instability (in the usual case of a
singly quantized vortex with $|q|=1$, this condition affects only
modes with $|m|=1$).

Rokhsar \cite{DR1} has
argued that a multiply quantized vortex is indeed unstable because of the
formation of a bound state of radius $\sim \xi$ inside the vortex core (for a
singly quantized vortex, he finds such a bound state  only for relatively
small condensates with $Na/d_0\lesssim 5$, where $d_0$ is the mean oscillator
length for the trap; a variational study of the Bogoliubov equations
confirms this qualitative conclusion).  In the TF limit,  the core radius
$\xi$ becomes small and  the details of the trap potential become
irrelevant for
phenomena on the scale of the vortex core, and the problem reduces to that of a
long vortex
  in an unbounded
condensate.  For this case, the radial eigenfunctions for normal modes that
are independent of $z$
do not decay
exponentially
 on the length scale $\xi$  \cite{ALF1}, precluding  such bound states as
solutions of
 the relevant Bogoliubov equations (see Sec.~V for a more detailed
discussion).

Although
the variational Eq.~(\ref{var2}) makes no use of the TF approximation,  it does
assume that the hydrodynamic variables provide the  appropriate
physical description.  For a uniform condensate in the Bogoliubov
approximation,  the hydrodynamic picture holds  only for
long-wavelength modes  with
$k\xi\lesssim 1$ when the
quasiparticle creation operator is an essentially equal admixture of a particle
and a hole creation operator \cite{FW,NP}. For $k\xi\gtrsim 1$, in contrast,
the quasiparticle creation operator becomes  a nearly pure particle
creation operator.

Correspondingly, a nonuniform trapped condensate has no hydrodynamic modes
unless it is sufficiently large  that
$\xi
\lesssim R_0$, since the system  otherwise approximates an ideal noninteracting
gas. Even for such a  large condensate,
however,  the only hydrodynamic modes are those  with
$\hbar\omega\lesssim \mu$, where $\mu$ is the chemical potential of the
condensate \cite{Str}.  This latter condition becomes progressively
more restrictive as
$N_0$ decreases and the condensate  approaches the ideal-gas limit.

\section{Perturbation analysis of the normal modes}

We return to the  hydrodynamic Eqs.~(\ref{hyd}).  In the present case of modes
$\propto e^{im\phi}$, they reduce to the coupled equations
\begin{mathletters}
\label{hydm}
\begin{eqnarray}
-i\bigg(\omega-\frac{mV}{\rho}\bigg) n' -\frac{m^2}{\rho^2}n_q\Phi'
+\bbox{\nabla\cdot}\big( n_q\,\bbox{\nabla}\Phi'\big)&=&0, \label{hyd3}\\
-i\bigg(\omega-\frac{mV}{\rho}\bigg)\Phi' +
\frac{g}{M} n'  +\frac{\hbar^2m^2}{4M^2n_q\rho^2}n'-
\frac{\hbar^2}{4M^2n_q}\bbox{\nabla\cdot}\left[n_q\bbox{\nabla} \left(\frac{
n'}{n_q}\right)\right] &=&0,\label{hyd4}
\end{eqnarray}
\end{mathletters}
where $n_q$ is the condensate density for the $q$-fold quantized vortex.

\subsection{Stationary condensate}

Any practical solution of these equations  requires   a physically motivated
approximation, and we here use the TF limit of a large condensate, when the
kinetic energy  in the GP equation (\ref{GP1}) is negligible compared to the
remaining terms.  In this limit, the equilibrium condensate density $n_0
=|\Psi|^2$ is given by
\begin{equation}
g|\Psi|^2 =\mu-\case{1}{2}Mv_0^2- V_{\rm tr},\label{TF}
\end{equation}
wherever the right-hand side is positive, and zero elsewhere;  the chemical
potential is determined by the normalization condition
\begin{equation}
N\approx N_0= \int dV\,|\Psi|^2 \label{norm}\end{equation}
(the present zero-temperature approximation neglects the difference between the
total number of particles  $N$ and the  number $N_0$ in the condensate).

 In the
simplest case of an axisymmetric trap potential (\ref{pot}) and a stationary
condensate with
${\bf v}_0=0$
\cite{BP}, Eq.~(\ref{TF}) gives the condensate density
\begin{equation}
gn_0 = \mu_0-V_{\rm tr},
\end{equation}
where $\mu_0$ is the chemical potential for a stationary condensate.  The
condensate density varies quadratically and  can be rewritten in terms
of the central density
$n_0(0)=\mu_0/g$
to give
\begin{equation}
\frac{n_0(\rho,z)}{n_0(0)} =
\bigg(1-\frac{\rho^2}{R_\perp^2}-\frac{z^2}{R_z^2}\bigg)\,
\Theta\bigg(1-\frac{\rho^2}
{R_\perp^2}-\frac{z^2}{R_z^2}\bigg)\label{n0}
\end{equation}
where $\Theta(x)$ is the unit positive step function. The condensate is
ellipsoidal with radial and axial dimensions
$R_\perp$ and $R_z$ given by
\begin{equation}R_\alpha^2 = \frac{2\mu_0}{M\omega_\alpha^2} =
\frac{2\mu_0}{\hbar\omega_\alpha}d_\alpha^2.\label{alpha1}\end{equation}
In the TF limit, each of these dimensions is much larger than the
corresponding oscillator lengths
$d_\alpha =
\sqrt{\hbar/M\omega_\alpha}$, so that $\mu_0\gg \hbar\omega_\alpha$.  Note that
the aspect ratio is given by
$R_z/R_\perp
= 1/\lambda$, with a  cigar-shaped (disk-shaped) condensate corresponding
to the
limit
$\lambda\ll 1$ ($\lambda\gg 1$).

The normalization condition (\ref{norm}) readily yields
\begin{equation}
\frac{Na}{d_0} =
\frac{1}{15}\bigg(\frac{2\mu_0}{\hbar\omega_0}\bigg)^{\!\!5/2}=
\frac{1}{15}\bigg(\frac{R_0}{d_0}\bigg)^{\!\!5}\gg 1,\label{radius}
\end{equation}
where $R_0 = (R_\perp^2R_z)^{1/3}$, $d_0 = (d_\perp^2d_z)^{1/3}$, and
$\omega_0=(\omega_\perp^2\omega_z)^{1/3}$ are appropriate ``geometric'' means.
 It is conventional to define the  coherence length $\xi$ in terms of the
central density:\begin{equation}
\xi^2= \frac{1}{8\pi n_0(0)a}.\end{equation}
  It satisfies the
simple relation
\begin{equation}\xi^2 = \frac{\hbar^2}{2\mu_0 M} =
\frac{d_0^4}{R_0^2},\label{xi}\end{equation}
implying  the set of TF inequalities $\xi \ll d_0\ll R_0$.

\subsection{Rotating condensate with a vortex}

In the presence of a quantized vortex with $q$ quanta of circulation on the
trap's symmetry axis,
  Eq.~(\ref{TF}) yields the corresponding condensate
density
\begin{equation}
gn_q = \mu_q - V_{\rm tr} - \case{1}{2}MV^2 ,
\end{equation}
where $\mu_q$ is the chemical potential for a $q$-fold  vortex and
$V=\hbar q/M\rho$.
This expression can be rewritten as
\begin{equation}
\frac{n_q(\rho,z)}{n_0(0)} =
\bigg(\frac{\mu_q}{\mu_0}-\frac{\rho^2}{R_\perp^2}-\frac{z^2}{R_z^2}-
\frac{q^2\xi^2}{ \rho^2}\bigg)\,
\Theta\bigg(\frac{\mu_q}{\mu_0}-\frac{\rho^2}
{R_\perp^2}-\frac{z^2}{R_z^2}-\frac{q^2\xi^2}{\rho^2}\bigg),
\label{nq}\end{equation}
using Eq.~(\ref{xi}).
The centrifugal barrier arising from the vortex alters the shape from
ellipsoidal  to toroidal; in the TF approximation, it  creates a flared
hollow core  with radius
$\approx|q|\xi$ in the plane $z=0$
\cite{DS,SS,DR1}.  The normalization condition (\ref{norm}) can be expanded to
show that the fractional change in the chemical potential
$(\mu_q-\mu_0)/\mu_0$  for fixed $V_{\rm tr}$ and $N$ is of order
$q^2d_0^4/R_0^4\,\ln\big(R_0^2/ |q|\,d_0^2\big)\ll 1$ \cite{SS}.

Equation (\ref{hyd4}) can be simplified considerably in the TF limit
when the kinetic energy is treated as small\cite{Str}.
 Specifically, Eq.~(\ref{hyd4}) contains the density fluctuation
$n'$ in two physically distinct ways. The last  two terms involving
$\hbar^2/M^2n_q$  reflect the quantum
kinetic-energy contribution to Bernoulli's equation;  in contrast,  the
preceding one involving $g/M$ reflects the  repulsive interparticle
interaction energy.  In the present case, the kinetic-energy  term for
$\rho\gtrsim \xi$ is smaller than the interparticle potential energy term  by a
factor of order
$\hbar^2/gMn_0(0)R_0^2\approx\xi^2/R_0^2=d_0^4/R_0^4$.  As
will be seen below, the effect of interest here is of order
$\xi/R_0=d_0^2/R_0^2$, so that the kinetic-energy term in Eq.~(\ref{hyd4})
can be
omitted entirely for
$\rho\gtrsim \xi$, giving
\begin{equation}
-i\bigg(\omega-\frac{m\hbar q}{M\rho^2}\bigg)\Phi' + \frac{g}{M} n'  =0.
\label{hyd5}
\end{equation}

A combination of Eqs.~(\ref{hyd3}) and (\ref{hyd5})
yields the single eigenvalue equation
\begin{equation}
\bbox{\nabla\cdot}\bigg(\frac {gn_q}{M}\bbox{\nabla} \Phi'\bigg)
-\frac{gn_q}{M}\,\frac{m^2}{\rho^2} \Phi'+
\bigg(\omega -
\frac{m\hbar
q}{M\rho^2}\bigg)^{\!\!2}\Phi' = 0
\end{equation}
This
second-order partial differential equation determines the frequencies and
normal
modes of an axisymmetric trapped condensate containing a vortex line through
terms of order
$d_\perp^2/R_\perp^2$. A combination of Eqs.~(\ref{alpha1})  and (\ref{radius})
shows that this small parameter has the following dependence
\begin{equation}
\frac{d_\perp^2}{R_\perp^2} = \left(\frac{d_\perp}{15
Na\lambda}\right)^{\!\! 2/5}\label{small}
\end{equation}
on the condensate
number $N$ and the asymmetry parameter $\lambda$

 It is now convenient to introduce dimensionless
units, with $\omega' \equiv
\omega/\omega_\perp$,
$\rho'
\equiv
\rho/R_\perp$, and $z' \equiv z/R_z = z\lambda/R_\perp$; in particular, the
dimensionless ratio
 $m\hbar q/\omega M\rho^2 = (m\,q\,d_\perp^2/R_\perp^2)\rho'^{-2}$ is
seen to be of order $d_\perp^2/R_\perp^2$ and thus small. In addition, the
actual condensate density $n_q$ differs from that for a vortex-free
condensate $n_0$ by corrections  of order $\xi^2/R_0^2 =
d_0^4/R_0^4$,
which is negligible compared to the effect of the superfluid flow.  Omitting
the primes on the dimensionless variables, we  therefore find
\begin{equation}
\frac{1}{2n_0(0)}\,\left[\frac{1}{\rho}\frac{\partial}{\partial
\rho}\left(\rho\,n_0\,\frac{\partial \Phi'}{\partial \rho}\right) -
\frac{m^2\,n_0\,\Phi'}{\rho^2} +
\lambda^2 \frac{\partial}{\partial
z}\left(n_0\frac{\partial \Phi'}{\partial z}\right)\right]+ \bigg(\omega^2-
\frac{2\omega\,m\,q}{\rho^2}\,\frac{d_\perp^2}{R_\perp^2}\bigg)
\Phi'= 0, \label{pert}\end{equation}
where   terms of order $d_\perp^4/R_\perp^4$ have been  omitted, and  the TF
condensate density $n_0$ is taken from Eq.~(\ref{n0}).

Equation (\ref{pert})  for the velocity potential $\Phi'$ contains
an explicit term of order $d_\perp^2/R_\perp^2$ arising from the circulating
condensate velocity. In fact, however, we seek  solutions  in a class of
functions for which the  particle-density fluctuation $n'$ is finite on
the vortex axis (at
$\rho=0$). To incorporate this boundary condition, it is natural rewrite
Eq. (\ref{pert}) in
terms of the function $n'$.  Equation~(\ref{hyd5}) shows that
$n'\propto \left(1-\frac{mqd_{\perp}^2}{\omega\rho^2R_{\perp}^2}\right)\Phi'$,
or equivalently,
$$\Phi'\propto
\left(1-\frac{mqd_{\perp}^2}{\omega\rho^2R_{\perp}^2}\right)^{-1}n'\approx
\left(1+\frac{mqd_{\perp}^2}{\omega\rho^2R_{\perp}^2}\right)\,n'. $$
Then, omitting terms of  order $d_{\perp}^4/R_{\perp}^4$,  we can replace
Eq.~(\ref{pert}) for $\Phi'$ by the following equation for $n'$:

\begin{equation}
\frac{1}{2n_0(0)}\,\left[\frac{1}{\rho}\frac{\partial}{\partial
\rho}\left(\rho\,n_0\,\frac{\partial n'}{\partial \rho}\right) -
\frac{m^2\,n_0\,n'}{\rho^2} +
\lambda^2 \frac{\partial}{\partial
z}\left(n_0\frac{\partial n'}{\partial z}\right)\right]+
\omega^2n'+\frac{2mqd^2_{\perp}}{\omega\rho^2R^2_{\perp}}
\bigg[1-\omega^2+\frac{n_0}{n_0(0)\rho}\left(\frac{1}{\rho}-\frac{\partial}
{\partial \rho}\right)\bigg]n'= 0\enspace ,
\label{pert1}
\end{equation}
where $n_0/n_0(0) = 1 -\rho^2-z^2$ in dimensionless units.

The normal modes of a large vortex-free ($q=0$) axisymmetric condensate
satisfy the following equation \cite{Fl,Ohb}
\begin{equation}
\frac{1}{2n_0(0)}\,\left[\frac{1}{\rho}\frac{\partial}{\partial
\rho}\left(\rho\,n_0\frac{\partial n'_0}{\partial \rho}\right) -
\frac{m^2\,n_0\,n'_0}{\rho^2} +
\lambda^2 \frac{\partial}{\partial
z}\left(n_0\,\frac{\partial n'_0}{\partial z}\right)\right]+
\left(\omega^0\right)^2
n'_0= 0, \label{pert0}
\end{equation}
where $\omega^0$ is the
frequency of a particular unperturbed normal mode. These low-lying TF normal
modes have been studied analytically for a spherical \cite{Str} and a
cylindrical
trap
\cite{Zar};  for a general axisymmetric trap, some of them  can be found
analytically
\cite{Str,Fl,Ohb}, but, in general, numerical techniques are necessary
\cite{HZ}.

Comparison of Eqs.~(\ref{pert1}) and (\ref{pert0}) immediately yields  a
simple perturbation expression for the frequency shift of any particular
normal
mode of an axisymmetric condensate induced by a
$q$-fold quantized vortex

\begin{equation}
\omega^2-\left(\omega^0\right)^2=
\frac{2\,m\,qd^2_{\perp}}{\omega^0R^2_{\perp}}
\left\langle\frac{(\omega^0)^2-1}{\rho^2}
+\frac{n_0}{n_0(0)\rho^3}\left(\frac{\partial}{\partial
\rho}-\frac{1}{\rho}\right)
\right\rangle\enspace,
\label{7}
\end{equation}
where the angular bracket denotes a matrix element evaluated with  the
appropriate  eigenfunction of the vortex-free condensate.  Note that this
result
describes only modes with $m\neq 0$ because the convective operator ${\bf
v}_0\cdot\bbox{\nabla} = (V/\rho)\partial/\partial \phi$ vanishes for an
axisymmetric state with $m = 0$.  To the accuracy considered here, the region
far from the vortex core makes the main contribution to the frequency shift,
and the result is independent of the detailed core structure.

One should mention that if we take  Eq.~(\ref{pert}) for  $\Phi'$, instead of
Eq.~(\ref{pert1}) for $n'$, as a basis for perturbation analysis, we obtain
equivalent results for the frequency shift only for $|m|\ge 2$. The point is
that the unperturbed solutions for
$\Phi'_0$ and $n'_0$ both behave like $\rho^{|m|}$ near the vortex line and, as
follows from Eq.~(\ref{hyd5}),  $n'$ will finite  at $\rho=0$ for such
solutions
only  if $|m|\ge 2$. But if we
treat  Eq. (\ref{pert1}) for the particle-density fluctuation $n'$
perturbatively, the boundary condition for $n'$ is
satisfied automatically for any value of $m$. As a specific example that
illustrates this distinction,
Eqs.~(\ref{dipole1}) and (\ref{dipole2}) show that the exact lowest
dipole modes of a vortex (those with
$|m|=1$)  have $n'\propto \rho$ and $\Phi'\propto \rho^{-1}$ for $\rho\to 0$.

\section{Frequency shifts for specific cases}

It is convenient to write the fractional shift in the squared frequency as
\begin{equation}
\frac{\omega^2-\left(\omega^0\right)^2}{\left(\omega^0\right)^2}\equiv q
\,\Delta\,{\rm sgn}\, m\,\frac{d_\perp^2}{R_\perp^2},
\label{shift}
\end{equation}
where ${\rm sgn}\,m = m/|m|$ and

\begin{equation}
\Delta=
\frac{2\,|m|}{(\omega^0)^3}
\left\langle\frac{(\omega^0)^2-1}{\rho^2}
+\frac{n_0}{n_0(0)\rho^3}\left(\frac{\partial}{\partial
\rho}-\frac{1}{\rho}\right)
\right\rangle\enspace .
\label{Delta}
\end{equation}
Since the matrix element is independent of the sign of $m$, the vortex,
generally speaking, splits the
two previously degenerate  modes with $\pm|m|$ into two  levels, according to
the sign of $m$. In particular, the (small) splitting for $|m|>0$ can be
characterized by the expression
\begin{equation}
\frac{\omega_{|m|}-\omega_{-|m|}}{\omega_{|m|}^0} \approx
q\,\Delta\,\frac{d_\perp^2}{R_\perp^2}=q\,\Delta\,
\left(\frac{d_\perp}{15Na\lambda}\right)^{\!\! 2/5},\label{split}
\end{equation}
where the last form follows from Eq.~(\ref{small}).
 The remaining calculation depends on the  shape of the
harmonic trap, and we consider several specific cases.

\subsection{Spherical trap}

For a spherical trap, each unperturbed eigenfunction is a product of  a
spherical harmonic and a radial polynomial~\cite{Str}
$$
n'_0\propto Y_{lm}(\theta,\phi)\,r^lP^{(l+\frac{1}{2},0)}_n(1-2r^2),
$$
where $P^{(l+\frac{1}{2},0)}_n(x)$ are Jacobi polynomials, $r^2=\rho^2+z^2$,
$\rho = r\sin\theta$, and
$n$ is the radial quantum number.
The corresponding dimensionless unperturbed eigenfrequency is given
by~\cite{Str} $\left(\omega_{nl}^0\right)^2 = l + n(2n+2l+3)$.

To calculate the matrix element, one  performs an angular and radial
integration. The angular averaging of the corresponding operators gives rise
to the
following result
\begin{equation}
\left\langle
\frac{1}{\sin^2\theta}
\right\rangle_\Omega=\frac{2l+1}{2|m|},
\end{equation}

\begin{equation}
\left\langle \frac{\cos\theta}{\sin^3\theta}\,\frac{\partial}{\partial
\theta}-\frac{1}{\sin^4\theta}
\right\rangle_\Omega=
\left\{\begin{array}{lcl}
-\frac{(2l+1)}{4|m|},\qquad |m|>1;
\\ \noalign{\vspace{0.2cm}}
\left.
-\frac{(2l+1)}{4}\left(1+\frac{l(l+1)}{2}\right),\qquad |m|=1.
\right.
\end{array}
\right.
\end{equation}

The remaining radial
integral can be evaluated using the following integrals for Jacobi polynomials
\cite{GR}
\begin{equation}
\int_{-1}^1
(1-x)^\alpha[P_n^{(\alpha,0)}(x)]^2=\frac{2^{\alpha+1}}{\alpha+2n+1}
\end{equation}
\begin{equation}
\int_{-1}^1 (1-x)^{\alpha-1}[P_n^{(\alpha,0)}(x)]^2=\frac{2^\alpha}{\alpha}
\end{equation}
\begin{equation}
\int_{-1}^1 (1-x)^{\alpha-2}[P_n^{(\alpha,0)}(x)]^2=\frac{2^{\alpha-1}}{\alpha}
\frac{(\alpha^2+2n\alpha+\alpha+2n^2+2n)}{(\alpha^2-1)}
\end{equation}
As a result, the general frequency shift for the spherical mode with quantum
numbers ($nlm$) is given  by the expression
\begin{equation}
{\Delta_{nlm}^{\rm sph}=
\left\{\begin{array}{lcl}
\frac{4n+2l+3}{\sqrt{l+n(2n+2l+3)}},\qquad |m|>1
\\
\left.
\frac{4n+2l+3}{\sqrt{l+n(2n+2l+3)}}-\frac{(4n+2l+3)l(l+1)
\left[(1+2n)(l+3/2)+2n^2\right]}
{4(l-1/2)(l+3/2)\left[l+n(2n+2l+3)\right]^{3/2}}
,\qquad |m|=1
\right.
\end{array}
\right.
}
\label{sph}
\end{equation}
Equations (\ref{split}) and (\ref{sph}) show that the vortex splits some (but
not all) of  the previously degenerate $2l+1$ modes with given ($nl$) and
different
$m$. The $m=0$ mode remains unshifted in this approximation, the $|m|=1$
mode is split by an amount $\Delta_{nl1}^{\rm sph}$, and  the remaining modes
with $2\le |m|\le l$  are split by a different amount $\Delta_{nlm}^{\rm sph}$
that is independent of the particular value of $m$.   For
$|m|=l=1$,
$n=0$ (the lowest dipole mode), one can  see
that the frequency shift  and splitting vanish, as expected, because the
vortex-free condensate and that with the vortex both  oscillate at the bare
trap frequency
$\omega_\perp$.

\subsection{Cylindrical trap}

The cylindrical case is obtained from the general trap potential by setting
$\omega_z=0$,   but it is also necessary to use the radius $R_\perp$ as
the characteristic length for both the radial and axial coordinates.  The
unperturbed condensate density is now given simply by $n_0(\rho)=
n_0(0)(1-\rho^2)$, and the perturbed eigenfunctions can be written in the
form $n'(\rho) e^{i(m\phi+kz)}$, where   $n'$ obeys the
eigenvalue equation
\begin{equation}
\left[\case{1}{2}(1-\rho^2)\left(\frac{1}{\rho}\frac{d}{d\rho}\rho\frac{d}{d\rho
}
-\frac{m^2}{\rho^2} -k^2R_\perp^2\right)-\rho\frac{d}{d\rho} +
\omega^2\right]n'(\rho)
+\frac{2mqd^2_{\perp}}{\omega\rho^2R^2_{\perp}}
\left[1-\omega^2+\frac{1-\rho^2}{\rho}\left(\frac{1}{\rho}-\frac{d}
{d \rho}\right)
\right]
n'(\rho) =0.
\end{equation}
The term with $k^2R_\perp^2$ precludes an exact solution \cite{Zar}, but the
frequency shift can again be found  perturbatively for wavelengths long
compared
to
$R_\perp$, when we find
\begin{equation}
\omega^2-\left(\omega^0\right)^2\approx
\frac{2\,m\,qd^2_{\perp}}{\omega^0R^2_{\perp}}
\left\langle\frac{(\omega^0)^2-1}{\rho^2}+\frac{1-\rho^2}{\rho^3}
\left(\frac{d}{d\rho}-\frac{1}{\rho}\right)
\right\rangle
+
\case{1}{2}k^2R_\perp^2\left\langle
1-\rho^2\right\rangle.
\end{equation}
Here, the angular brackets denote  a matrix element evaluated with the
unperturbed eigenfunctions, which have the explicit form
$n'_0\propto \rho^{|m|}P_n^{(|m|,0)}(1-2\rho^2)$,
and the corresponding eigenfrequency is
given by \cite{Fl,Zar} $\left(\omega_{nm}^0\right)^2 = |m|+ 2n(n+|m|+1)$.

The fractional shift in the squared eigenfrequencies is readily evaluated with
the same radial integrals used for the spherical case
\begin{equation}
\frac{\omega_{nm}^2-(\omega_{nm}^0)^2}{(\omega_{nm}^0)^2}=
\frac{k^2R_\perp^2}{2(|m|+2n)(|m|+2n+2)}
+q\,{\rm sgn}\,m\, \frac{d_\perp^2}{R_\perp^2}\times
\left\{\begin{array}{lcl}
\frac{(4n+2|m|+2)}{\sqrt{|m|+2n(n+|m|+1)}}, \qquad |m|>1
\\ \noalign{\vspace{0.2cm}}
\left.
\frac{4n(n+1)(n+2)}{\left[1+2n(n+2)\right]^{3/2}},\qquad |m|=1
\right.
\end{array}
\right.
\enspace,
\label{cyl}
\end{equation}
where the second term is the contribution of the vortex and the first is that
of the traveling wave.
The vortex again has only a small effect, which implies
that the configuration is stable, assuming that perturbation theory is valid.
For $|m|=1$, $n=0$, $k=0$ (dipole mode), the frequency shift is equal to zero.

\subsection{General axisymmetric trap}
The normal modes of a general axisymmetric trap in the TF limit have been
classified completely  and reduced to the diagonalization of  a sequence of
increasingly large matrices
\cite{Fl,Ohb}.  We shall use the cylindrical
coordinates of Ref.~\cite{Ohb}, which turn out to be more   convenient
for the present purpose. The unnormalized solutions of Eq.~(\ref{pert0})
have the
form
$n'^{+}_0(\rho,z) =
\rho^{|m|}B_{nm}^+$ and $n'^{-}_0(\rho,z)=\rho^{|m|}zB_{nm}^-$
and are, respectively, even
and odd under the transformation $z\to -z$ (our notation differs somewhat from
that of Ref.~\cite{Ohb}).  Here,
$B_{nm}^+$ and
$B_{nm}^-$ are even polynomials with typical terms $\rho^{2n_1}z^{2n_2}$,
where
$n_1$, $n_2$, and $n$ are  non-negative integers with $n_1+n_2\le
n$.  For any given
$n$ and
$m$, there are
$n+1$ even-parity  normal modes and, separately, $n+1$
odd-parity  normal modes, labeled by $j =  0, 1, \cdots, n$,  with frequencies
$\omega_{jnm}^{0\pm}$ determined by diagonalizing
$(n+1)$-dimensional matrices
 given  in Ref.~\cite{Ohb}.  For $n=0$, the even
($+$) and odd ($-$) modes have dimensionless frequencies  $\omega^{0+}_{j0m} =
\sqrt{|m|}$ and
$\omega^{0-}_{j0m} =
\sqrt{|m|+\lambda^2}$, respectively, with $B_{0m}^+ = B_{0m}^- = 1$ and $j=0$.
For general values of the asymmetry parameter $\lambda =
\omega_z/\omega_\perp$,
it is straightforward to determine the corresponding solutions for $n = 1$
\cite{Ohb}, but the higher-order modes require numerical analysis (for
$\lambda\to 0$, Ref.~\cite{Fl} shows that all the unperturbed frequencies agree
with those for a cylindrical trap).

Equations (\ref{shift}) and (\ref{Delta}) determine the  vortex-induced
frequency shift $\Delta_{jnm}^\pm$ of any particular normal mode. The matrix
element should be evaluated for $j = 0, 1, 2, \cdots, n$ (the $n+1$
distinct  eigenfunctions for a given $nm\pm$).  For the simplest case of $n=0$,
we find

\begin{equation}
\Delta_{00m}^+=
\left\{\begin{array}{lcl}
  \frac{2|m|+3}{\sqrt{|m|}}
,\qquad |m|>1
\\ \noalign{\vspace{0.2cm}}
\left.
0,\qquad |m|=1
\right.
\end{array}
\right.
\end{equation}
\begin{equation}
\Delta_{00m}^-=
\left\{\begin{array}{lcl}
\frac{2|m|+5}{\sqrt{|m|+\lambda^2}}
,\qquad |m|>1
\\ \noalign{\vspace{0.2cm}}
\left.
\frac{7\lambda^2}{(1+\lambda^2)^{3/2}},\qquad |m|=1\enspace .
\right.
\end{array}
\right.
\end{equation}
These expressions  are  equivalent to those found in Ref.~\cite{SS}
apart from the case  $|m|=1$, where our treatment of the singular terms
gives additional terms,  ensuring that the lowest dipole mode is unshifted
because
$\Delta^+_{001}=0$.

As a check, we note
that
$\Delta_{00m}^+$ is just
$\Delta_{nlm}^{\rm sph}$  from Eq.~(\ref{sph}) for a spherical condensate with
$n=0$ and $l =|m|$, and that $\Delta_{00m}^-$ for $\lambda=1$ is just
$\Delta_{nlm}^{\rm sph}$ for a
spherical condensate with $n=0$ and $l =|m|+1$. In addition,
Eq.~(\ref{split}) with our values for
$\Delta_{002}^+= 7/\sqrt 2$ and
$\Delta_{001}^-=7\lambda^2/(1+\lambda^2)^{3/2}$ 	yields frequency
splittings
that agree with those found independently from sum rules by Zambelli and
Stringari~\cite{ZS} for the lowest quadrupole modes with $|m|=2$ and $|m|=1$
in the TF limit.

 For the next  case of $n=1$, Figs.~\ref{fig1} and
\ref{fig2}  show the corresponding dimensionless  eigenfrequencies
$\omega_{j1m}^{0\pm}$ and  fractional frequency shifts
and splittings $\Delta_{j1m}^{\pm}$ for the even and odd modes
$j = 0,1$  and $|m| = 1, 2$ as a function of the asymmetry parameter
$\lambda$ at
fixed $d_\perp^2/R_\perp^2$. It is interesting that the curves for
$\Delta_{011}^{\pm}$ and $\Delta_{111}^{\pm}$ are non-monotonic functions of
$\lambda$ and reach a maximum near $\lambda=1$, which is a spherical trap (this
behavior is also seen in $\Delta^-_{001}$ above).
  In contrast to the situation for the unperturbed  frequencies, the frequency
shifts
$\Delta^\pm$ for a long cigar-shaped   axisymmetric trap (the limit $\lambda\to
0$) differ from those for a cylindrical trap because of the different angular
integrations.
 To check  these results, the particular case of
$\lambda=1$ (a spherical trap) was studied analytically for $n=1$, confirming
that these normal modes are indeed consistent with those found in
Ref.~\cite{Str}.

\section{Bogoliubov equations and vortex stability}

In the Bogoliubov approximation, the ``grand-canonical hamiltonian''
operator $\hat K=\hat H - \mu\hat N$ reduces to a pure condensate part $K_0$
that depends only on the condensate wave function
$\Psi$, and a noncondensate part $\hat K'$ \cite{AP,ALF}.  The latter
 involves  the solutions of the Bogoliubov equations,
\begin{eqnarray}
{\cal L}u_j-g\Psi^2v_j&=&E_ju_j,\\
{\cal L}v_j-g\left(\Psi^*\right)^2u_j&=&-E_jv_j,
\end{eqnarray}
where ${\cal L} = T+V_{\rm tr} + 2g|\Psi|^2-\mu$.  Here, $u_j$ and
$v_j$
 are a complete set of
coupled  amplitudes
 that obey the normalization condition $\int
dV\,\left(|u_j|^2-|v_j|^2\right) = 1$, and $E_j$ is the associated eigenvalue.
The ground-state occupation of the
$j$th mode is simply $N_j'=\int dV |v_j|^2$, and the noncondensate part
(a second-quantized  operator) reduces to
\begin{equation}
\hat K' = -{\sum_j}'E_jN_j' + {\sum_j}'E_j\beta_j^\dagger\beta_j,\label{nonc}
\end{equation}
where the primed sum means to omit the
condensate mode, and $\beta_j^\dagger$ and $\beta_j$ are quasiparticle
creation and annihilation operators that obey boson commutation relations
$[\beta_j,\beta_k^\dagger]=\delta_{jk}$.  The ground state of $\hat K'$ is
the vacuum $|{\bf 0}\rangle$, defined by $\beta_j|{\bf 0}\rangle = 0$ for all
$j\neq 0$, so that the first term in Eq.~(\ref{nonc}) is the ground-state
expectation value of $\hat K'$;  the remaining
 term  involves the quasiparticle number operator
$\beta_j^\dagger\beta_j$, and the spectrum of $\hat K'$ is bounded from
below as long as $E_j>0$ for all $j$.  If, however, any of the properly
normalized solutions has a negative eigenvalue, the system can lower its
ground-state thermodynamic potential $\Omega = \langle \hat K\rangle$
arbitrarily
by creating more and more quasiparticles for that particular normal mode. This
behavior  indicates that the original ground state $\Psi$ is unstable and
must be
reconstructed to form a new stable ground state \cite{BD}.

\subsection{Axial vortex waves}

For a condensate  containing a $q$-fold quantized vortex in an axisymmetric
trap (studied in Sec.~II), the condensate wave function has the form
$\Psi({\bf r}) = e^{iq\phi}|\Psi_q(\rho,z)|$. In the presence of a
general condensate velocity ${\bf v}_0 = (\hbar/M)\bbox{\nabla} S$, it is
convenient to
make explicit the phase of the condensate wave function through
factors $e^{\pm iS}$ \cite{ALF},  and the Bogoliubov amplitudes here can be
written as
\begin{equation}
\pmatrix{u({\bf r}) \cr v({\bf
r})}=e^{im\phi+ikz}\,\pmatrix{e^{iq\phi}\,\tilde u_m(\rho, z)\cr
e^{-iq\phi}\,\tilde v_m(\rho,z)}.\label{uvm}
\end{equation}
This form represents a helical wave
 with wavenumber $k$ and angular momentum $m$ relative to the
condensate; we assume that the  factor $e^{ikz}$ varies rapidly compared to
the amplitudes
$\tilde u$ and $\tilde v$, with the condensate containing many wavelengths
($kR_z\gg 1$).
The corresponding amplitudes obey the coupled equations
\cite{LPP}
\begin{mathletters}
\label{Pit}
\begin{eqnarray}
\left[-\frac{\hbar^2}{2M}\nabla^2
+\frac{\hbar^2}{2M}
\left(\frac{(m+q)^2}{\rho^2}+k^2-2ik\frac{\partial}{\partial z}\right)+V_{\rm
tr}+2g|\Psi_q|^2-\mu\right]\tilde u_m-g|\Psi_q|^2\tilde v_m&=&E\tilde u_m,\\
\left[-\frac{\hbar^2}{2M}\nabla^2
+\frac{\hbar^2}{2M}
\left(\frac{(m-q)^2}{\rho^2}+k^2-2ik\frac{\partial}{\partial z}\right)+V_{\rm
tr}+2g|\Psi_q|^2-\mu\right]\tilde v_m-g|\Psi_q|^2\tilde u_m&=&-E\tilde v_m,
\end{eqnarray}
\end{mathletters}
where $\nabla^2= \partial^2/\partial\rho^2+\rho^{-1}\partial/\partial \rho
+\partial^2/\partial z^2$ is expressed in cylindrical polar coordinates.
The different centrifugal barriers imply that the two amplitudes behave
differently near the axis of symmetry, with $\tilde u_m \propto \rho^{|m+q|} $
and
$\tilde v_m\propto
\rho^{|m-q|}$ as $\rho\to 0$.

As noted in Sec.~II, these Bogoliubov amplitudes are linearly related to the
corresponding hydrodynamic variables $n'$ and $\Phi'$.  In an axisymmetric
trap with a
$q$-fold vortex on the symmetry axis, their angular dependence is simply $
e^{im\phi}$, and the corresponding amplitudes are
given by Eqs.~(\ref{n'Phi'})
\cite{ALF,WG,FR}
\begin{mathletters}
\label{hydamp}
\begin{equation}
n_{jm}'= |\Psi_q|\left(\tilde  u_{jm}-\tilde v_{jm}\right),
\end{equation}
\begin{equation}
\Phi_{jm}' = \frac{\hbar}{2Mi\,|\Psi_q|}\left(\tilde  u_{jm}+\tilde
v_{jm}\right),
\end{equation}
\end{mathletters}
where $j$ denotes the remaining set of quantum numbers, and $|\Psi_q|\propto
\rho^{|q|}$ for $\rho\to 0$.  In the simplest case of a vortex-free condensate
($q=0$), these amplitudes reproduce the behavior $\propto \rho^{|m|}$
found in Refs.~\cite{Fl,Ohb}, but the situation is more complicated for
$q\neq0$.  The boundary condition that  $\tilde u$ and $\tilde v$ remain
bounded near the origin (discussed below in Sec.~V.B and in Ref.~\cite{ALF1})
 implies a corresponding behavior for
$n'$ and
$\Phi'$.  Although
$\Phi'$ has an apparent   singularity for small
$\rho$ from the factor
$|\Psi_q|^{-1}$,   physical quantities like the
  hydrodynamic current involve an additional
factor
$n_q = |\Psi_q|^2$
\cite{ALF}.  As shown in Sec.~III, the detailed core structure has negligible
effect on the normal-mode eigenfrequencies in the TF limit (a large
condensate),
but it would be significant for a smaller condensate with
$Na/d_0
\lesssim 10$, when the TF relation in Eq.~(\ref{radius}) implies that
$d_0/R_0\gtrsim 0.4$ is no longer small \cite{DBEC}.

The general  Bogoliubov equations are difficult to solve. In the
present case of a harmonic trap potential $V_{\rm tr}$, however, they have
three  exact solutions.  For these special ``dipole'' modes,  any condensate
described by a wave function
$\Psi$ that satisfies the GP equation (\ref{GP1}) undergoes a
rigid oscillation along the three principal axes
of the trap  with the bare oscillator frequencies \cite{FR}. Define the
raising and lowering operators
\begin{mathletters}
\label{eq:aadag}
\begin{equation}
a_\alpha^\dagger \equiv
\frac{1}{\sqrt{2}} \bigg(\frac{x_\alpha}{d_\alpha} -
d_\alpha{\partial \over {\partial x_\alpha}}\bigg),
\label{eq:adag}
\end{equation}
\begin{equation}
a_\alpha \equiv
\frac{1}{\sqrt{2}} \bigg(\frac{x_\alpha}{d_\alpha} +
d_\alpha{\partial \over {\partial x_\alpha}}\bigg),
\label{eq:a}
\end{equation}
\end{mathletters}
where  $d_\alpha \equiv \sqrt{\hbar/m\omega_\alpha}$, and $\alpha = x$, $y$,
and $z$.
  As shown in Ref.~\cite{FR}, the state
\begin{equation}
{\cal U}_\alpha({\bf r})  \equiv\pmatrix{u_\alpha({\bf r})\cr
 v_\alpha({\bf r})}
=
\pmatrix{a_\alpha^\dagger \Psi({\bf r})\cr a_\alpha
\Psi^*({\bf r})}
\label{eq:uv}
\end{equation}
satisfies the coupled Bogoliubov equations with an eigenvalue
$E_\alpha=\hbar\omega_\alpha$.

For an axisymmetric trap potential $V_{\rm tr} =
\frac{1}{2}M\omega_\perp^2\rho^2
+ \frac{1}{2}M\omega_z^2z^2$, the two states ${\cal U}_x$ and ${\cal U}_y$ are
degenerate, each with energy $\hbar\omega_\perp$.  In terms of  the new
operators
\cite{CT}

\begin{eqnarray}
a_\pm = \frac{1}{\sqrt 2}\left(a_x\mp ia_y\right),\\
a_\pm^\dagger = \frac{1}{\sqrt 2}\left(a_x^\dagger\pm i a_y^\dagger\right),
\end{eqnarray}
the (conserved) angular momentum about the symmetry axis can be rewritten as
$L_z = \hbar(a_+^\dagger a_+-a_-^\dagger a_-)$.  As a result,  $a_+^\dagger$
($a_-^\dagger$) adds a unit of positive (negative) angular momentum.
Correspondingly, we have two distinct solutions of the Bogoliubov equations
\begin{equation}
{\cal U}_\pm({\bf r}) =\pmatrix{ u_\pm({\bf r})\cr\noalign{\vspace{.1cm}}
 v_\pm({\bf r})}=  {\cal U}_x({\bf r}) \pm i \, {\cal U}_y({\bf r})
=
\pmatrix{a_\pm^\dagger\Psi({\bf r})\cr\noalign{\vspace{.1cm}} a_\mp\Psi^*({\bf
r})},
\end{equation}
each with energy $\hbar\omega_\perp$.  In  cylindrical polar coordinates, the
explicit form of these amplitudes is readily determined;  for a $q$-fold
quantized vortex,  we find
\begin{mathletters}
\label{plus}
\begin{eqnarray}
 u_+({\bf r})  &=&
\case{1}{2}e^{i(q+1)\phi}\left(\frac{\rho}{d_\perp}
-d_\perp\frac{\partial}{\partial
\rho} +
\frac{q d_\perp}{\rho}\right)|\Psi_q(\rho, z)|,\\ \noalign{\vspace{.1cm}}
 v_+({\bf r})  &=& \case{1}{2}e^{i(-q+1)\phi}\left(\frac{\rho}{d_\perp}
+d_\perp\frac{\partial}{\partial
\rho} +
\frac{q d_\perp}{\rho}\right)|\Psi_q(\rho, z)|,
\end{eqnarray}
\end{mathletters}
\begin{mathletters}
\label{minus}
\begin{eqnarray}
 u_-({\bf r})  &=&
\case{1}{2}e^{i(q-1)\phi}\left(\frac{\rho}{d_\perp}
-d_\perp\frac{\partial}{\partial
\rho} -
\frac{q d_\perp}{\rho}\right)|\Psi_q(\rho, z)|,\\ \noalign{\vspace{.1cm}}
 v_-({\bf r})  &=& \case{1}{2}e^{i(-q-1)\phi}\left(\frac{\rho}{d_\perp}
+d_\perp\frac{\partial}{\partial
\rho} -
\frac{q d_\perp}{\rho}\right)|\Psi_q(\rho, z)|,
\end{eqnarray}
\end{mathletters}
and it is clear that these solutions are properly orthogonal, with $\int
dV\,\left( u_-^* u_+- v_-^* v_+\right)=0$.  Comparison
with Eq.~(\ref{uvm}) shows that
$ {\cal U}_\pm$ has
$m = \pm1$ and
$k=0$.  Pitaevskii
\cite{LPP} constructed analogous zero-frequency solutions in the case of a
vortex
line in an unbounded fluid (his solutions emerge in the limit $\omega_\perp\to
0$, so that $d_\perp\to\infty$).

It is worth  noting that the fluctuations in the particle density $n'$
and the velocity potential $\Phi'$ for these lowest dipole-mode solutions
(\ref{plus})-(\ref{minus}) have a simple explicit expressions:
\begin{equation}
n'=-\frac{d_{\perp}}{2}\,e^{ \pm i\phi}\,\frac{\partial n_q}{\partial
\rho},\label{dipole1}
\end{equation}
\begin{equation}
\Phi'=\frac{\hbar}{2Mi}\,e^{\pm
i\phi}\left(\frac{\rho}{d_{\perp}}\pm\frac{qd_{\perp}}{\rho}\right)
\,.\label{dipole2}
\end{equation}
The contrasting behavior of $n'$ and $\Phi'$ near the vortex axis helps
motivate the discussion in Sec.~III.B.

The form of Eq.~(\ref{Pit}) suggests an approximation for small $k$ and
$m=\pm1$, using $ {\cal U}_\pm$ as the unperturbed eigenfunctions.
Standard first-order perturbation theory gives the corrected eigenvalue
\begin{equation}
E_{\pm }(k)\approx \hbar\omega_\perp + \frac{\hbar^2k^2}{2M}\,\frac{\int
dV\left(| u_\pm|^2+| v_\pm|^2\right)}{\int
dV\left(| u_\pm|^2-| v_\pm|^2\right)},
\end{equation}
where the additional term involving $\partial/\partial z$ yields a surface
contribution that vanishes.  Use of Eqs.~(\ref{plus}) and (\ref{minus}) yields
the explicit expressions
 \begin{mathletters}
\begin{equation}
\int
dV\left(| u_\pm|^2+| v_\pm|^2\right) = 2\int
dV\,\left[\left(\frac{\rho}{d_\perp}\pm
q\frac{d_\perp}{\rho}\right)^{\!\!2}|\Psi_q|^2+ d_\perp^2
\left(\frac{\partial|\Psi_q|}{\partial\rho}\right)^{\!\! 2}\right],\label{num}
\end{equation}
\begin{equation}
\int
dV\left(| u_\pm|^2-| v_\pm|^2\right) = 2\int dV\,\left(\rho\pm
q\frac{d_\perp^2}{\rho}\right)\left(-\frac{\partial
|\Psi_q|^2}{\partial\rho}\right).\label{den}
\end{equation}
\end{mathletters}
Equation (\ref{den}) is easily evaluated with an integration by parts and
the normalization condition Eq.~(\ref{norm}).  In the TF limit of a large
condensate with $d_\perp\ll R_\perp$, the dominant contribution to
Eq.~(\ref{num})  comes far from the vortex core, so that we can replace
$|\Psi_q|$ from Eq.~(\ref{nq}) by
$|\Psi_0|$ from Eq.~(\ref{n0}).  Neglecting corrections of order
$d_\perp^2/R_\perp^2$, we find the compact result
\begin{equation}
E_\pm(k) \approx\hbar\omega_\perp +
\frac{\hbar^2k^2}{2M}\,\left(\frac{R_\perp^2}{7d_\perp^2}\pm q\right)
=
\hbar\omega_\perp\left[1+\case{1}{2}\left(\case{1}{7}k^2R_\perp^2\pm
q\,k^2d_\perp^2\right)\right].
\end{equation}
As in Eq.~(\ref{cyl}), the correction is indeed small if  $|q|$ is not too
large and $kR_\perp\lesssim 1$  (since we also require $kR_z\gg 1$,
this latter condition implies a cigar-shaped condensate with small
asymmetry parameter $\lambda
\ll 1$). Consequently, a
$q$-fold quantized vortex has long-wavelength  propagating axial helical modes
 with wavenumber $k$ and angular distortion
$m=\pm 1$.  In this perturbative limit, the
frequencies  of   these two modes exceed the bare dipole frequency
$\omega_\perp$, as expected for an internal distortion.  The
separate parabolic branches  are  split by an amount
\begin{equation}
E_+(k)-E_-(k) \approx q\,\frac{\hbar^2k^2}{M}.\end{equation}

In his original work on a singly quantized vortex in an
unbounded dilute Bose gas, Pitaevskii \cite{LPP} developed
an approximate description of
vortex  waves for shorter
wavelengths, assuming only that $k\xi\ll 1$.  To logarithmic
accuracy,  the dispersion relation has the classical form
$\omega_k=\left(\hbar k^2/2M\right)\ln\left(1/k\xi\right)$~\cite{Thom}. This
method is easily generalized to the present case of a $q$-fold quantized vortex
on the symmetry axis of an axisymmetric trap in the TF limit.  We use the
coherence length $\xi$ as the length scale and  rewrite the exact condensate
density as $\psi_q^2 \equiv |\Psi_q|^2/n_0(0)$.  The coupled Bogoliubov
equations (\ref{Pit}) for
$m = -1$ become (here, the coordinates are dimensionless)
\begin{mathletters}
\label{Pit1}
\begin{eqnarray}
\left[-\nabla^2 +\frac{(q-1)^2}{\rho^2} +\kappa^2
-2i\kappa\frac{\partial}{\partial z} + \frac{\xi^4}{d_\perp^4}\left(\rho^2 +
\lambda^2z^2\right) + 2\psi_q^2-1\right]\tilde u_{-1} -\psi_q^2\tilde v_{-1}
&=&
\epsilon \tilde u_{-1},\\
\left[-\nabla^2 +\frac{(q+1)^2}{\rho^2} +\kappa^2
-2i\kappa\frac{\partial}{\partial z} + \frac{\xi^4}{d_\perp^4}\left(\rho^2 +
\lambda^2z^2\right) + 2\psi_q^2-1\right]\tilde v_{-1} -\psi_q^2\tilde u_{-1}
&=& -\epsilon \tilde v_{-1},
\end{eqnarray}
\end{mathletters}
where we have set $\mu_q\approx \mu_0 = gn_0(0)$ and introduced the
abbreviations $\epsilon = (2E/\hbar\omega_\perp)(\xi^2/d_\perp^2)=E/\mu_0 \ll
1$
 and $\kappa = k\xi \ll 1$.

The amplitudes vary slowly along the axis with a
characteristic length $R_z$, so that the various derivatives with respect
to $z$
can be omitted, along with the trap potential (which is here of order
$\xi^4/d_\perp^4\ll 1$), in which case these
equations (\ref{Pit1}) take the approximate form
 \begin{mathletters}
\label{Pit2}
\begin{eqnarray}
\left(-\frac{1}{\rho}\frac{d}{d\rho}\rho\frac{d }{d
\rho} +\frac{(q-1)^2}{\rho^2} +\kappa^2 +
 2\psi_q^2-1\right)\tilde u_{-1} -\psi_q^2\tilde v_{-1} &=&
\epsilon \tilde u_{-1} ,\\
\left(-\frac{1}{\rho}\frac{d}{d\rho}\rho\frac{d }{d
\rho} +\frac{(q+1)^2}{\rho^2} +\kappa^2 +
 2\psi_q^2-1\right)\tilde v_{-1} -\psi_q^2\tilde u_{-1} &=&
-\epsilon \tilde v_{-1}.
\end{eqnarray}
\end{mathletters}
In the same approximation, the condensate wave function satisfies the following
GP equation
\begin{equation}
\left(-\frac{1}{\rho}\frac{d}{d\rho}\rho\frac{d }{d
\rho} +\frac{q^2}{\rho^2} +
 \psi_q^2-1\right)\psi_q= 0,
\end{equation}
whose asymptotic solution for $\rho\gg 1$ is $\psi_q \approx 1-q^2/2\rho^2$.
This set of equations is now precisely the same as that obtained by Pitaevskii
\cite{LPP}, apart from the appearance of the quantum number $q$ instead of
$q=1$.  Exactly the same arguments then show that the dispersion relation is
given by
$E_k\approx q(\hbar^2k^2/2M)\ln(1/k\xi)$, where we assume $k\xi \ll 1\ll
kd_\perp$.  This condition implies that $E_k$ is significantly greater than the
lowest dipole energy $E_\perp =\hbar\omega_\perp= \hbar^2/Md_\perp^2$.

\subsection{Bound core states}

Rokhsar \cite{DR1} has argued that the mere existence of a bound state
localized
in the vortex core implies an instability, with the original axisymmetric
configuration undergoing a displacive transition of the vortex core.  In the
present context, such an instability would arise only from the presence of one
or more negative-energy eigenvalues of the Bogoliubov equations, causing  the
original ground state of Eq.~(\ref{nonc}) to collapse.  For such a
situation, our
preceding perturbative approach would  fail completely, requiring a more
powerful
approach.

In their original solution for a long vortex line in  unbounded
superfluid $^4$He, Ginzburg and Pitaevskii \cite{GP} remarked that multiply
quantized vortices are energetically unfavorable. It is also notable
that no clear evidence for  doubly quantized vortices in superfluid ${}^4$He
II has ever
been reported \cite{RJD}. Thus, we
consider only the case $q=1$  and seek solutions with relative angular symmetry
$\propto e^{im\phi}$ for the full Bogoliubov equations with a singly
quantized vortex in an axisymmetric trap.  In the TF limit,
we neglect small corrections of order $\xi^2/R_0^2\sim\xi^4/d_0^4$, and these
equations then reduce to those for a vortex line in an unbounded condensate
[here, they are direct generalizations of Eqs.~(\ref{Pit2}), obtained for
general $m$  by setting $q=1$]. For $q=1$, the result of Eq.~(\ref{var2})
shows that the only possible unstable solution corresponds to the choice
$|m|=1$. Then, since the amplitude
$\tilde u_{-1}$ remains
finite at the origin, we anticipate that the choice  $m=-1$  will yield a
positive normalization for $|\tilde u_{-1}|^2-|\tilde v_{-1}|^2$.

These coupled Bogoliubov radial
equations (\ref{Pit2}) are equivalent to a single fourth-order ordinary
differential equation.  Near
the origin, there are four linearly independent solutions, and it is
straightforward to show
\cite{ALF1} that only two are finite as $\rho\to 0$.  In a matrix notation,
they can be written
$(\tilde u_{-1}^{(1)},\tilde v_{-1}^{(1)})$ and
$(\tilde u_{-1}^{(2)},\tilde v_{-1}^{(2)})$.  Each is a power series in
$\rho^2$,  with leading behavior (omitting constant factors)
 \begin{equation}
\pmatrix{\tilde u^{(1)}(\rho)\cr
\noalign{\vspace{.1cm}}
\tilde
v^{(1)}(\rho)\cr}\approx\pmatrix{\rho^0\cr\noalign{\vspace{.1cm}}\rho^4\cr}
\qquad\hbox{and}\qquad
\pmatrix{\tilde u^{(2)}(\rho)\cr
\noalign{\vspace{.1cm}}
\tilde
v^{(2)}(\rho)\cr}\approx\pmatrix{\rho^6\cr\noalign{\vspace{.1cm}}\rho^2\cr},
\end{equation}
where we have suppressed the subscript $-1$.  The general solution is a linear
combination, with $\tilde u\approx c_1 +{\cal O}(\rho^2)$ and $\tilde v\approx
c_2\rho^2 +{\cal O}(\rho^4)$, where $c_1$ and $c_2$ are constants. Note that
the linear combinations $\tilde u\pm\tilde v$ appearing in the
hydrodynamic amplitudes [Eqs.~(\ref{hydamp})] both have the same behavior at
the origin.  Although this  situation apparently produces a singularity in the
variational expression in Eq.~(\ref{var2}), it is not hard to verify that the
divergence in fact cancels out.

Similarly, it is not difficult to verify \cite{ALF1} that the asymptotic
solutions of Eqs.~(\ref{Pit2}) have the form $\propto
\rho^{-1/2}\exp(i\zeta\rho)$, where $\zeta^2$ is the solution of a quadratic
equation.  One solution is negative with $\zeta^2 = -\zeta_{\rm im}^2 = -
\sqrt{\epsilon^2+1}-\kappa^2-1$; the other is $\zeta_1^2 =
\sqrt{\epsilon^2+1}-\kappa^2-1$, which can take either sign.  To eliminate the
growing exponential solution $\propto \rho^{-1/2}\exp(|\zeta_{\rm im}|\rho)$,
it is necessary to adjust the ratio $c_1/c_2$, leaving a solution for
$\tilde u$ and $\tilde v$ that is bounded both at the origin and at infinity.
It has the asymptotic form
\begin{equation}
\tilde u, \tilde v\sim
\rho^{-1/2}\big[A\exp(i\zeta_1\rho)+B\exp(-i\zeta_1\rho)+ C\exp(-|\zeta_{\rm
im}|\rho)\big],
\end{equation}
where $A$, $B$, and $C$ depend on $\epsilon$ and $\kappa$.

For $\kappa^2 >0$, it is easy to see that
$\zeta_1$ is real if
$|\epsilon|>\epsilon_B(\kappa)\equiv
\sqrt{\kappa^4 + 2\kappa^2}$, which is
the Bogoliubov energy for the wave number $\kappa$
along the vortex axis. For this range of parameters, it follows
that the asymptotic  solutions oscillate and there is no bound
state.
 In contrast, if
$|\epsilon|<\epsilon_B(\kappa)$, then $\zeta_1$
becomes imaginary, and the coefficient of the growing
exponential must vanish.  If we choose Im $\zeta_1>0$, then
$B(\kappa,\epsilon) $ must vanish, which in principle
determines the dispersion relation of the bound state as a
function of the axial wave number.   In
effect, this procedure yields  Pitaevskii's  vortex-wave solution
\cite{LPP}.

The situation is different if $\kappa = 0$, for
then $\epsilon_B(0)$ vanishes, and $\zeta_1^2
=
\sqrt{\epsilon^2+1}-1$ is non-negative for all
real $\epsilon$.  Thus the general solution for
$\kappa=0$  necessarily oscillates far from the
vortex core for any real positive  value of $\epsilon^2$,
implying that there is no bound state in this
case (the solution with $\epsilon=0$ is simply the dipole
oscillation, here  with zero frequency, as found by
Pitaevskii \cite{LPP}).  A similar argument holds
  in the extreme TF limit for general $q$ and $m$, since there
are always two linearly independent solutions that remain finite
at the origin \cite{ALF1}, and the asymptotic behavior for large $\rho$ is
independent of the values of $q$ and $m$ (the centrifugal
barriers contribute only in higher orders).

This behavior is easily understood from a classical perspective.  A long
classical hollow-core vortex line in an incompressible fluid has
irrotational flow that is describable with a velocity
potential. In addition to boundary conditions at the inner surface of the
core, the velocity potential satisfies Laplace's equation \cite{ALF2}.
Solutions of Laplace's equation cannot exhibit spatial oscillations in all
directions, so that a vortex wave propagating along the axis  with
wavenumber
$k$ necessarily decays exponentially  $\sim e^{-k\rho}$ in the radial
direction.  In the limit $k\to 0$, this exponential decay length
$k^{-1}$ diverges, and the amplitude can only vanish algebraically as $\rho\to
\infty$, which is the behavior found here for a vortex line in an unbounded
condensate.

       Based on Rokhsar's discussion \cite{DR1},  the numerical study of Dodd
{\it et al.\/} \cite{DBEC} found one negative-frequency normal-mode  solution
for a singly quantized vortex (the actual negative frequency should be the
negative of that shown in their Fig.~2).  In the ideal-gas limit, the
anomalous  frequency is $-\omega_\perp$,  and it increases toward zero from
below in the large-condensate limit $N_0a/d_0\gg 1$.  The present treatment
implies that  this particular eigenfrequency approaches zero from below in the
limit of a large condensate ($\xi/R_0\ll 1$).  The precise dependence on the
small parameter, and the relation to the critical angular velocity for vortex
creation remains for future analysis.

\section{Discussion}

We have used a variety of methods to study the stability of an axisymmetric
vortex in a large  axisymmetric trapped Bose condensate.   In this limit,
the repulsive interparticle condensate energy dominates the condensate kinetic
energy.  The resulting Thomas-Fermi (TF) approximation holds when the
equatorial
condensate radius $R_\perp$ is much larger than the corresponding oscillator
length $d_\perp = \sqrt{\hbar/M\omega_\perp}$. The hydrodynamic description
of the condensate's normal modes simplifies greatly in the TF limit, and the
presence of vortex leads to small corrections of relative order
$d_\perp^2/R_\perp^2\ll 1$.  A perturbation analysis shows that a vortex
splits
the degenerate normal modes of the vortex-free condensate (those for $\pm m$)
by an amount
 of this same order, which should provide a clear signal for the existence of
a vortex in the trapped condensate.  Since the vortex-free condensate is
considered stable from both a theoretical and an experimental perspective
\cite{Edw,Str,EDCB,Rup,Jin,MOM}, this conclusion indicates that the vortex
should also be stable, at least in perturbation theory.   To preclude the
possibility of nonperturbative effects associated with the possible presence of
bound states with a radial  range $\sim \xi$ localized in the vortex core
\cite{DR1}, we  demonstrate that a long vortex line in an unbounded condensate
(the extreme TF limit of a trapped condensate) has no such normal modes
\cite{ALF1} other than the vortex waves found by Pitaevskii \cite{LPP}.

\acknowledgments

We are grateful for many stimulating discussions with
D.~Rokhsar and S.~Stringari.  This work was supported in part
by the National Science Foundation, under Grant No.~DMR 94-21888 and by  a
Stanford Graduate Fellowship (AAS).

\begin{figure}
\caption{The  $\lambda$-dependence (asymmetry parameter)  of the
unperturbed dimensionless frequency
$\omega_{j1m}^{0+}$ and the fractional  frequency shift $\Delta_{j1m}^+$
for  even modes with $n = 1$ and
$j = 0, 1$  (note that $\lambda\ll 1$ represents a cigar-shaped condensate).
Solid lines denote
$m=1$ and dashed lines denote $m=2$. }
\label{fig1}
\end{figure}

\begin{figure}
\caption{The  $\lambda$-dependence (asymmetry parameter)  of the
unperturbed dimensionless frequency
$\omega_{j1m}^{0-}$ and the fractional  frequency shift $\Delta_{j1m}^-$
for  odd modes with $n = 1$ and
$j = 0, 1$  (note that $\lambda\ll 1$ represents a cigar-shaped condensate).
Solid lines denote
$m=1$ and dashed lines denote $m=2$. }
\label{fig2}
\end{figure}

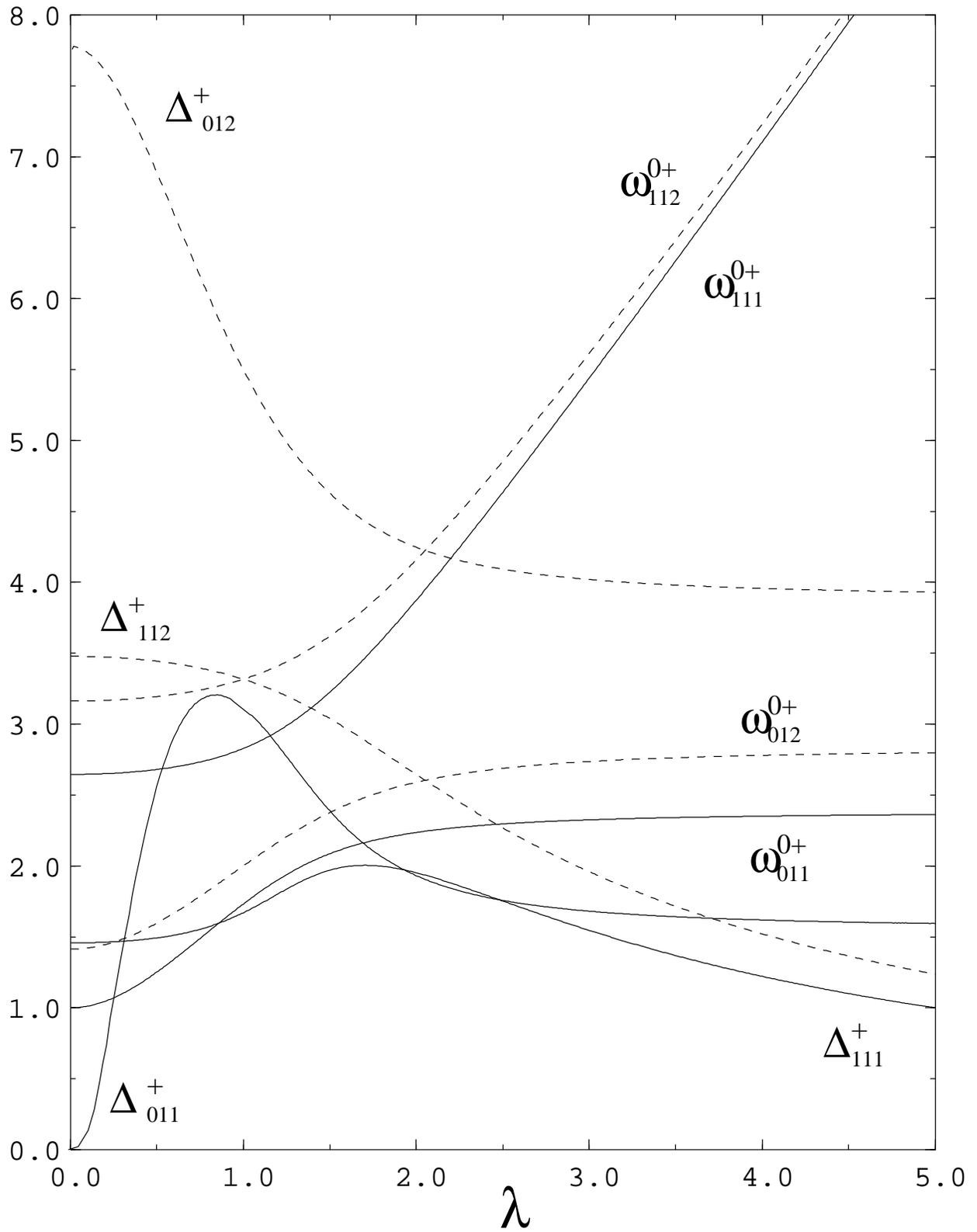
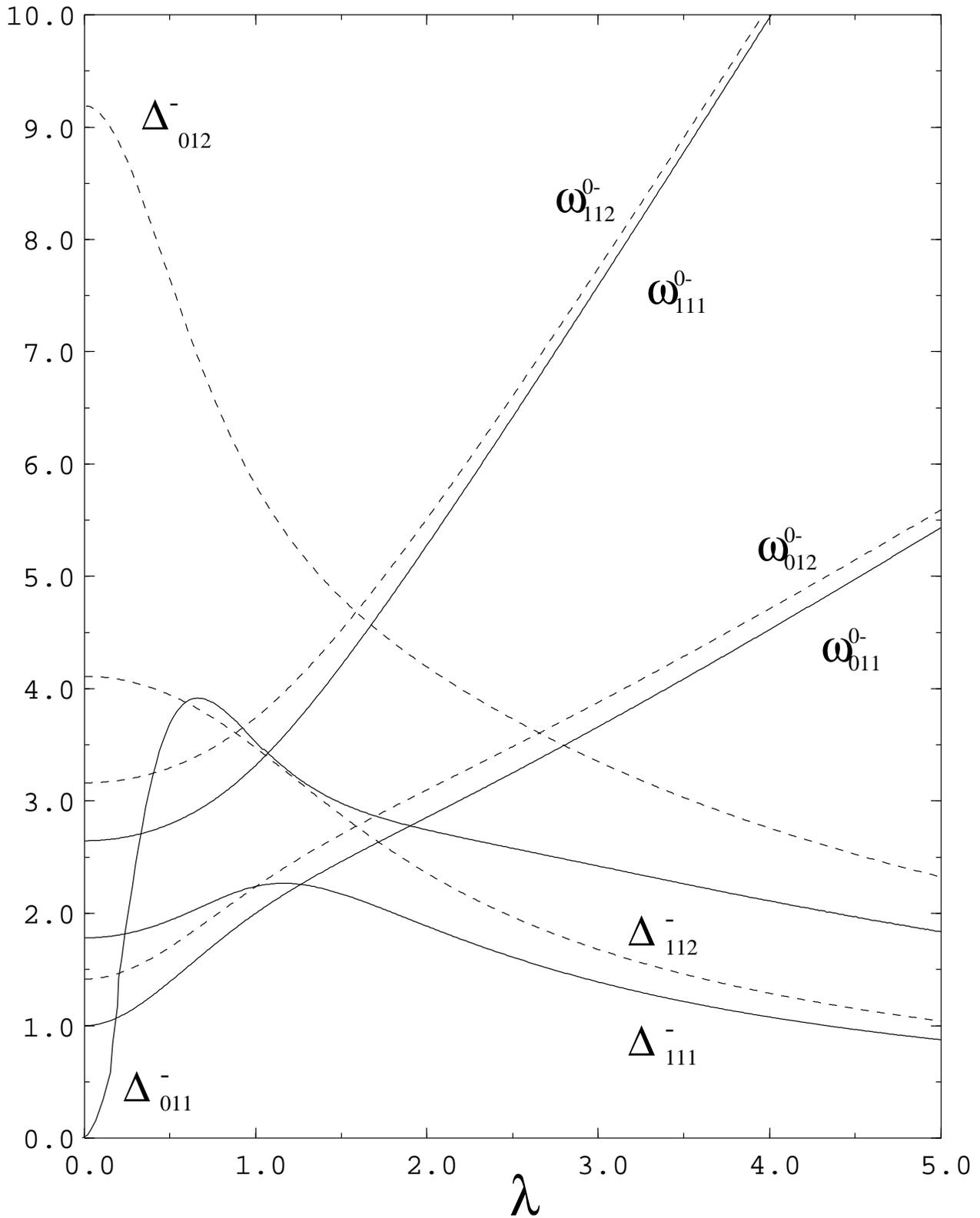
\end{document}